\documentclass[%
 aip,
 amsmath,amssymb,
 reprint,%
]{revtex4-1}

\usepackage{graphicx}
\usepackage{dcolumn}
\usepackage{bm}

\usepackage[utf8]{inputenc}
\usepackage[T1]{fontenc}
\usepackage{mathptmx}
\usepackage{etoolbox}
\usepackage[normalem]{ulem}

\newcommand{\blue}{\textcolor{black}}

\usepackage[version=4]{mhchem}
\usepackage{color}
\makeatletter
\def\@email#1#2{%
 \endgroup
 \patchcmd{\titleblock@produce}
  {\frontmatter@RRAPformat}
  {\frontmatter@RRAPformat{\produce@RRAP{*#1\href{mailto:#2}{#2}}}\frontmatter@RRAPformat}
  {}{}
}%
\makeatother
\begin{document}

\title[4-pol.]{
Determination of Stokes vector from a single image acquisition  
}

\author{Shuji Kamegaki}
\affiliation{CREST-JST and School of Materials and Chemical Technology, Tokyo Institute of Technology,
Ookayama, Meguro-ku, Tokyo 152-8550, Japan}
\author{Meguya Ryu}
\affiliation{National Metrology Institute of Japan (NMIJ), National Institute of Advanced Industrial Science and Technology (AIST), Tsukuba Central 3, 1-1-1 Umezono, Tsukuba 305-8563, Japan}
\author{Soon Hock Ng}
\affiliation{Optical Sciences Centre, ARC Training Centre in Surface Engineering for Advanced Materials (SEAM), 
Swinburne University of Technology, Hawthorn, Victoria 3122, Australia}
\author{Vygantas Mizeikis}
\affiliation{Research Institute of Electronics, Department of Electronics and Materials Science, Graduate School of Engineering, Shizuoka University, 3-5-1 Johoku Naka-ku, Hamamatsu 432–8011, Japan}
\author{Sean Blamires}
\affiliation{Mark Wainwright Analytical Centre, School of Biological, Earth and Environmental Science, University of New South Wales, Sydney NSW 2052, Australia}
\affiliation{School of Mechanical and Mechatronic Engineering, University of Technology, Sydney NSW 2007, Australia}
\author{Saulius Juodkazis}
\affiliation{Optical Sciences Centre, ARC Training Centre in Surface Engineering for Advanced Materials (SEAM), 
Swinburne University of Technology, Hawthorn, Victoria 3122, Australia}
\affiliation{WRH Program International Research Frontiers Initiative (IRFI) Tokyo Institute of Technology, Nagatsuta-cho, Midori-ku, Yokohama, Kanagawa 226-8503 Japan}
\affiliation{Laser Research Center, Physics Faculty, Vilnius University, Saul\.{e}tekio Ave. 10, 10223 Vilnius, Lithuania}
\author{Junko Morikawa}
\affiliation{WRH Program International Research Frontiers Initiative (IRFI) Tokyo Institute of Technology, Nagatsuta-cho, Midori-ku, Yokohama, Kanagawa 226-8503 Japan}
\affiliation{CREST-JST and School of Materials and Chemical Technology, Tokyo Institute of Technology,
Ookayama, Meguro-ku, Tokyo 152-8550, Japan}

\date{\today}%

\begin{abstract}
Four Stokes parameters (1852) define the polarisation state of light. Measured changes of the Stokes vector of light traversing an inhomogeneous sample are linked to the local anisotropies of absorption and refraction and are harnessed over an increasing range of applications in photonics, material, and space/earth observation. Several independent polarisation sensitive measurements are usually required for determination of the all four Stokes parameters, which makes such characterisation procedure time-consuming or requires complex setups. Here we introduce a single-snapshot approach to Stokes polarimetry in transmission by use of a 4-polarisation camera with the on-chip integrated polarisers. A quarter-waveplate was added in front of the sample and was illuminated by a linearly polarised light. This approach is demonstrated by measuring birefringence $\Delta n\sim 0.012$ of  spider silk of only $\sim 6~\mu$m-diameter using microscopy, however,  due to its generic nature, it is transferable to other spectral ranges and imaging applications, e.g., imaging from a fast moving satellite or drone or monitoring fast changing events such as phase transitions.    
\end{abstract}

\keywords{Anisotropy, polarisation analysis, Stokes parameters, polarimetry }

\maketitle

\section{\label{intro}Introduction}

Orientation is an important property of materials, underlying the optical~\cite{adom.202101004}, mechanical~\cite{Yamada}, and thermal~\cite{xu2019nanostructured} behaviours. Imaging of aligned structures in biological samples~\cite{Bueno,Bone:07,Sobczak:21} and tissues are linked to various medical conditions and is of the highest interest for the development of new medical instrumentation and bio-scaffolds for tissue regeneration~\cite{Yasuno}. An optical 3D tomography based on \blue{holography is demonstrated for imaging of a 3D micro-object which has combined birefringence and absorption inhomogenieties~\cite{Saba:21}. Recently, a 3D tomography of an object was extracted from optical images using a tensorial description of light-matter interaction~\cite{Park}. These microscopic techniques are expected to find a wide range of applications in microscopy.} \blue{3D tomography can even be made} with non-propagating optical near-fields, using polarisation-controlled input and polarisation analysis at the output for the far-IR and IR wavelengths in a popular attenuated total reflection geometry~\cite{22nh1047}. Orientation can be revealed from transmission or reflection even under conditions where feature sizes of $\sim 100$~nm of an aligned pattern are $\sim 20$ times below the diffraction limit, as shown for the IR chemical fingerprinting wavelengths where nanoscale resolution can not be reached~\cite{19n732}. Mechanical stress-induced anisotropies can be imaged by change of characteristic wavenumbers in Raman spectroscopy/scattering as demonstrated for sapphire~\cite{Pezzotti}. Peculiarities of population dynamics of photoexcited carriers inside semiconductors can be revealed by the transient grating method with polarisation sensitive probing or excitation~\cite{Garmire}.  Apart of science based methods, polarisation analysis is widely adopted in industry for edge and alignment detection~\cite{Otani},  satellite polarisation imaging~\cite{ZHANG200221}, an inspection of stress distribution in semiconductors and solar cells~\cite{Takeuchi}. \blue{Stokes polarimetry, which determines the variation of the polarisation state across the image of an optical vortex~\cite{citeme1} and measurement of the polarisation state at separate RGB colors in the image~\cite{citeme2} can be carried out using integrated polarisers and a quarter waveplate. Integration of polarisation elements even to a higher degree of complexity is demonstrated with metamaterials, where the state of polarisation, including a waveplate function can be realized on a flat surface in front of detector~\cite{Capa}.  } 

Here, we 
introduce a fast single snap-shot polarimetry technique based on the four polarisation (4-pol.) imaging. It was developed for absorption anisotropy  (dihroism) mapping~\cite{Hikima} and now it is demonstrated for birefringent objects (optical retarders) using a 4-pol. camera with an on-chip integrated wire-grid polarisers. Determination of all four Stokes parameters, which fully describe the polarisation state of propagating light, was carried out in a single image acquisition using 
a liquid crystal retarder with a $\lambda/4$ waveplate retardace inserted before the sample. The amplitude of the retardance and its orientation azimuth were both retrieved from a single measurement. This opens the feasibility for real-time imaging and monitoring of anisotropy changes during phase transitions, e.g.,  amorphous-to-crystalline changes.

\setcounter{figure}{0}


\section{Samples and Methods }\label{exp}

We used a $2^{12}$-level grey scale CMOS camera (CS505MUP1 Thorlabs) for imaging in the visible spectral range. It has an extinction ratio larger than 100 for the entire visible spectral range with maximum at $Ext\equiv \frac{T_{max}}{T_{min}}\sim 400$. 
While the intensity which is defined by the sum of all four images appears as usual, the orientation of linear polarisers is revealed by calculating the azimuth angle  $\theta_{shift} = \frac{1}{2}\arctan_2\left(\frac{I_{45} - I_{-45}}{I_{0} - I_{90}} \right)$ for each pixel, where $\arctan_2$ is the 2-argument arc-tangent $\theta = \arctan_2(y,x)$ with $-\pi<\theta\leq\pi$, i.e., it returns angle $\theta$ in the full $[-\pi,\pi]$ range~\cite{21ep46}. The images in this study 
were calculated from original four images without relying of the intensity and azimuth images provided by the software which controls the camera. 

\begin{figure*}[t]
\centerline{\includegraphics[width=0.8\linewidth]{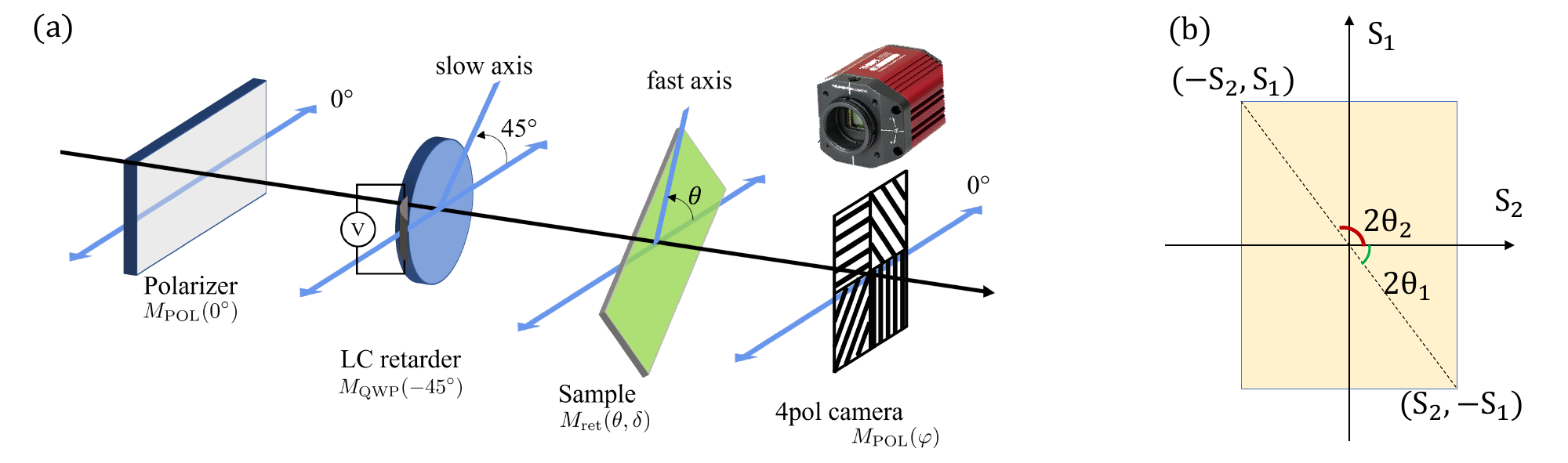}}
\caption{(a) Polarisation analysis with four-polarisation (4-pol.) camera (CS505MUP1 Thorlabs). Polarisers are on-chip integrated and the pixel size is $3.45~\mu$m. The anti-clockwise direction when looking into the beam corresponds to positive $\varphi$ angles. The liquid crystal retarder at $-45^\circ$ defines its fast axis. (b) Conventions for the Stokes parameters $S_{1,2}$ and azimuth angles $\theta_{1,2}$ used in analysis.  The ambiguity of slow vs. fast axis definition is due to $\tan 2\theta = -S_1/S_2$ which has solutions $2\theta = 2\theta -\pi$ (or $\theta = \theta-\pi/2$, i.e., the slow or fast axis marked by the dashed line diagonal is not determined unequivocally). A simple cross-Nicol image with a 530-nm color shifter $\lambda$-plate can determine slow vs. fast orientation~\cite{23lpr2200535}.} \label{f-meth}
\end{figure*}

For determination of the retardance $\delta$ and its orientation azimuth $\theta$ we used a broadband $350-700$~nm liquid crystal retarder at the $\lambda/4$ waveplate setting (LCC1223T-A Thorlabs) and a narrow-band $\sim 20$~nm filter to select illumination wavelength, e.g., $550\pm 10$~nm (Fig.~\ref{f-meth}).

Silk from golden orb web spiders \emph{Trichonephila plumipes} was used in this study; they were collected during the night 
in Sydney, Australia. Their major ampullate (dragline) silk was collected by forcible spooling~\cite{Blamires2}. 
For silk harvesting, spiders were anesthetized using \ce{CO2}, placed ventral side up on a foam platform, and 
fixed using non-adhesive tape and pins. Then, a single silk thread from the spinnerets was collected under a dissecting microscope and an electronic spool rotating at 1~m/min was used to reel silk threads from spiders~\cite{Blamires1}. 

\begin{figure*}[t]
\centerline{\includegraphics[width=.85\linewidth]{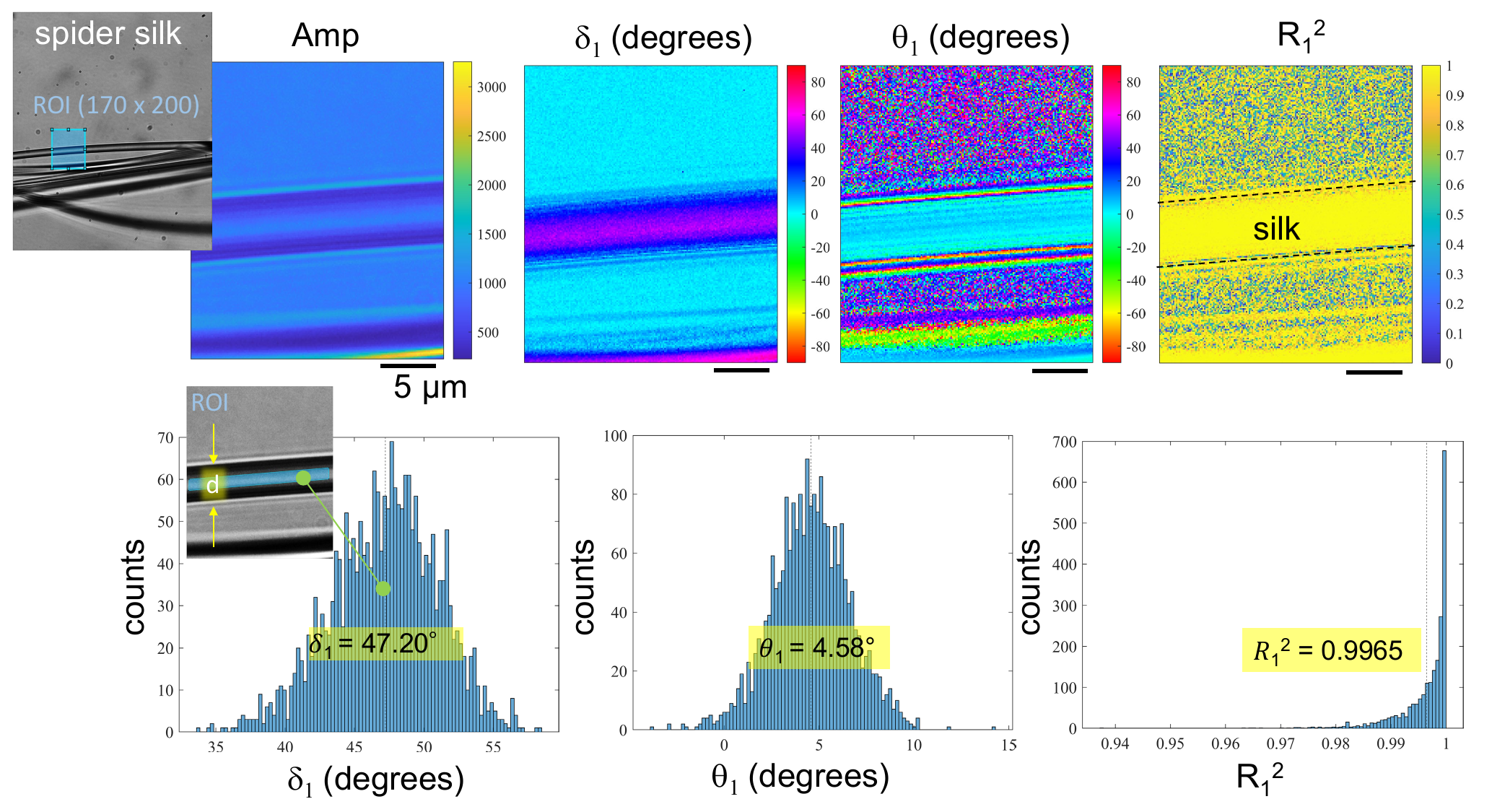}}
\caption{Polarisation analysis of yellow spider silk \emph{Trichonephila plumipes}. 
Single pixel fit of 4-pol. images (from region of interest ROI $170\times 200$ pixels) by Eqn.~\ref{e-s0} for 4-angles $\varphi$: $Fit(\varphi) = Amp\times\left[(1 - \sin{\delta}\sin(2(\varphi-\theta))/2\right]$ is plotted for amplitude $Amp$, retardance $\delta$, its azimuth $\theta$ and residuals $R^2$ between the fit and experimental values. Statistical distribution of $\delta$ and $\theta$ and $R^2$ from the central region of spider silk fiber are presented in histograms. 
Diameter of the silk fiber was $d = 6~\mu$m (the largest outside cross section between the intensity maxima; see Fig.~\ref{f-cros}). Objective lens $NA = 0.65$ (Nikon) was used for imaging at the corresponding resolution $r = 0.61\lambda/NA \approx0.52~\mu$m ($\lambda=550~$nm). Since $\sin\delta = \sin(\pi-\delta)$ for $0\leq\delta\leq\pi$ (I\&II quadrants), there are two possible values for $\delta_1$
; also, $\sin\delta=\sin{(-\pi-\delta)}$ for $-\pi\leq\delta\leq 0$ (IV\&III quadrants). This analysis is valid for the limits of $0^\circ\leq \delta\leq 90^\circ$ and $-90^\circ\leq \theta\leq 90^\circ$; see, Fig.~\ref{f-comp} for the complementary possible case of the fit.} \label{f-map}
\end{figure*}

\section{Theory: image of retardation and its azimuth}\label{Theory}

\subsection{Best fit based analysis}

Imaging with a 4-pol. camera can show the retardance distribution and azimuth of a retarder by numerical processing of 4 images as shown in this section. The experimental set up for measurements of the three Stokes parameters $S_{0,1,2}$ in transmission consist of a linear polarizer, sample (an unknown retarder) and 4-pol. camera. However, an additional quarter wave plate is required for the determination of retardance vector (retardance and its azimuth). The fourth Stokes parameter $S_3$ can be determined after such modification of setup as shown in this study. 


 The Mueller matrix formalism was used to describe the experiment. The polarizer at an angle of $\varphi$ is given by Muller matrix $M_{pol}$:
 \begin{equation}
\mathbf{M}_{pol}(\varphi)= 
    \begin{pmatrix}
      1 &  \cos2\varphi &  \sin2\varphi &  0 \\
      \cos2\varphi &  \cos^22\varphi & \cos2\varphi\sin2\varphi &  0 \\
      \sin2\varphi & \cos2\varphi\sin2\varphi &  \sin^22\varphi  & 0 \\
      0 & 0 & 0 & 0
    \end{pmatrix}.
 \end{equation}
The matrix can be set to correspond to the wire grid polarizers on the pixels of the CMOS detector. Angles $\varphi$ are set on each quadrant of the camera at $\varphi = 0,\pi/4,\pi/2,3\pi/4$, respectively. The sample acting as a retarder plate with the phase retardance of $\delta$ and retardance azimuth of $\theta$ is given by:
\begin{widetext}
\begin{equation}\label{ret}
\mathbf{M}_{ret}(\theta,\delta)= 
    \begin{pmatrix}
      1 &  0 &  0 &  0 \\
      0 & \cos^22\theta+\cos\delta\sin^22\theta  &  (1-\cos\delta)\sin2\theta\cos2\theta &  \sin\delta\sin2\theta \\
       0 &  (1-\cos\delta)\sin2\theta\cos2\theta &  \sin^22\theta+\cos\delta\cos^22\theta & -\sin\delta\cos2\theta \\
      0 & - \sin\delta\sin2\theta & \sin\delta\cos2\theta & \cos\delta
    \end{pmatrix}.
\end{equation}
\end{widetext}
\blue{This description (Eqn.~\ref{ret}) is for the pure birefringent material and does not take into account dichroism or optical activity. This is a limitation of this approach, which is most useful for birefringence dominant cases of imaged objects.  }

A quarter-wave plate is added for the determination of the phase retardance $\delta$ and retardance azimuth $\theta$. The quarter-wave plate aligned along the $-\pi/4$ direction is given by (follows from Eqn.~\ref{ret}):
\begin{equation}\label{lambda4}
\mathbf{M}_{\lambda/4}(-\pi/4)= 
    \begin{pmatrix}
        1 & 0 & 0 & 0 \\
        0 & 0 & 0 & -1 \\
        0 & 0 & 1 & 0 \\
        0 & 1 & 0 & 0 \\
    \end{pmatrix}
\end{equation}

The incident light was set to the horizontal (x-direction) so that it can be described by the Stokes vector $\mathbf{S}^{IN} = (1,1,0,0)$. Finally, the output light (Stokes vector) detected by the CMOS 4-pol. camera is the solution of the following matrix equation: 
\begin{align}\label{e-mat}
 \mathbf{S}^{OUT}(\varphi,\theta,\delta) &= \mathbf{M}_{pol}(\varphi)\cdot\mathbf{M}_{ret}(\theta,\delta)\cdot\mathbf{M}_{\lambda/4}(-\pi/4)\cdot\mathbf{S}^{IN} \\
 &= \frac{1}{2}
    \begin{pmatrix}
        1-\sin{\delta}\sin{2(\varphi-\theta)} \\
        \cos{2\varphi}\left(1-\sin{\delta}\sin{2(\varphi-\theta)}\right) \\
        \sin{2\varphi}\left(1-\sin{\delta}\sin{2(\varphi-\theta)}\right) \\
        0
    \end{pmatrix}
\end{align}
The Eqn.~\ref{e-mat} defines the all four output Stokes parameters $\mathbf{S}^{OUT}$ ($S_3 =0$ due to linear polarisers on 4-pol. camera segments). 

Next, we  use the first $S_0$, which is intensity, since we measure intensity images in experiment. It has this form, which is a result of a simple matrix multiplication:

\begin{equation}\label{e-s0}
    \mathbf{S}^{OUT}_0(\varphi,\theta,\delta) = \frac{1}{2} - \frac{1}{2}\sin{\delta}\sin{2(\varphi-\theta)}.
\end{equation}
By fitting intensity ($S_0$) at four angles $\varphi$, the amplitude, retardance $\delta$ and its azimuth $\theta$ can be determined. This protocol was used to determine the retardance and its azimuth $\delta,\theta$ of yellow spider silk fibers, which are examples of a birefringent (retarder waveplate) material at visible wavelengths. By virtue of simultaneous acquisition of four intensity images at different polarisations and addition of a $\lambda/4$-plate to the setup, the retardance and its orientation can be obtained from the single measurement. The quality of the best fit was assessed by calculating $R^2$ residuals for experimentally determined intensity $I_{i}$ of each pixel in the image by  (index \blue{$i=1-4$} corresponds to angles $\varphi = 0,\pi/4,\pi/2,3\pi/4$):
\begin{equation}
    R^2 = 1-\frac{\Sigma_{i=1}^4(I_i - f(i))^2}{\Sigma_{i=1}^4(I_i - \overline{I}^2)},
\end{equation}
\noindent where the fit function is $f(\varphi) = Amp\times \left[1 - \sin{\delta}\sin{2(\varphi-\theta)}\right]/2$ (see, Eqn.~\ref{e-s0}). The fit is obtained for two pairs of $(\delta_1,\theta_1)$ for $0\leq\delta\leq\pi/2$ and $(\delta_2,\theta_2)$ for $-\pi/2\leq\delta\leq 0$ corresponding to the positive and negative birefringence, respectively. Regardless of $\delta$ sign, the values are calculated from the same image and $\delta_2 = -\delta_1$, $\theta_2 = \theta_1 -\pi/2$. The direction of fast vs. slow axis, $\theta_1$ or $\theta_2$, is determined unequivocally from the cross-Nicol imaging with complimentary color shifting $530$~nm $\lambda$-plate, see, e.g., for the polyhydroxybutyrate (PHB) bounded spherulite organic crystal~\cite{23lpr2200535}.   

\subsection{Analytical solution based analysis}

The Stokes vector after passing the $\lambda/4$-plate at the setting of right-hand circular (RHC; the fast axis of liquid crystal $\lambda/4$ plate is at $-45^\circ$) polariser and sample (retarder plate; Fig.~\ref{f-meth}) is: 
\begin{equation}\label{e-asol}
S' = M_{ret}(\delta,\theta)\cdot S_{RHC} =
\begin{pmatrix}
        1 \\
        \sin{2\varphi}\sin{\delta}\\
        -\cos{2\varphi}\sin{\delta} \\
        \cos{\delta}
    \end{pmatrix}
\end{equation}
Experimentally detected intensity by 4-pol. camera is (polariser at $\varphi$ angles):
\begin{equation}
S_0(\varphi)\equiv I_\varphi=\frac{1}{2}(S'_0+S'_1\cos2\varphi + S'_2\sin2\varphi).
\end{equation}
For the 4-pol. angles $\varphi$, the corresponding intensities reads: $I_0 = (S'_0 + S'_1)/2$, $I_{90} = (S'_0-S'_1)/2$, $I_{45} = (S'_0+S'_2)/2$, and $I_{-45} = (S'_0-S'_2)/2$. The Stokes parameters $S'$ are directly related to the measured intensity by the 4-pol. camera: $S'_0 =(I_{0} + I_{45} + I_{90} + I_{-45})/2$, $S'_1 = I_0 - I_{90}$, $S'_2 = I_{45} - I_{-45}$. There is an ambiguity with $S'_3$ parameter since two angles $\theta_{1,2}$ satisfy $\tan 2\theta = -S'_1/S'_2$: $\theta_1 = \arctan_2(-S'_1,S'_2)/2$ and $\theta_2 = \arctan_2(S'_1,-S'_2)/2$ (Fig.~\ref{f-meth}(b) and Fig.~\ref{f-atan}). The corresponding retardances $\delta_{1,2}$, which are expressed via Stokes parameters $S'_{1,2}$:   $\delta_1 = \arcsin{(S'_1/\sin2\theta_1)} = \arcsin{(-\sqrt{S\mathrm{'}_{1}^2 + S\mathrm{'}_{2}^2})}$ and $\delta_2 = \arcsin{(S'_1/\sin2\theta_2)} =  \arcsin{(\sqrt{S\mathrm{'}_{1}^2 + S\mathrm{'}_{2}^2})}$. Set of $(\delta_{1,2}, \theta_{1,2})$ satisfies $S'_3(\delta_1) = S'_3(\delta_2)$ and can be calculated using experimentally determined 4-pol. intensity images for $S'_{0,1,2}$ as described above. 

\section{Results} 

\subsection{Retardance and its azimuth $\delta,\theta$ from fit of 4-pol. images}

Measurement of orientational properties of an absorbing sample requires only one linear polariser for the incoming light onto the sample or at the detector. With a 4-pol. camera, orientational dependence of absorbance can be easily determined with a single measurement~\cite{Hikima}. 
For samples which have distributed birefringence, a retarder plate plus two polarisers are required and can be used in crossed or aligned geometries with the sample rotated between them~\cite{19nh1443}. From the angular dependence of transmittance, the absorption and birefringence contributions are separated due to the double modulation frequencies of absorbance, $\varphi$ and $2\varphi$, respectively~\cite{19ass127}. The birefringence mapping requires a larger number of measurements due to a faster oscillation of transmittance vs. polarisation angle. As shown in Sec.~\ref{Theory}, it is possible to extract retardance and its orientations from the generic expression Eqn.~\ref{e-s0} when the additional $\lambda/4$ plate is introduced before the sample. For the orientation of retardance, an optical activity based polariscopy method has been developed for anisotropy mapping using true color coding of the orientation~\cite{Michael}.

\begin{figure}[tb]
\centerline{\includegraphics[width=1\linewidth]{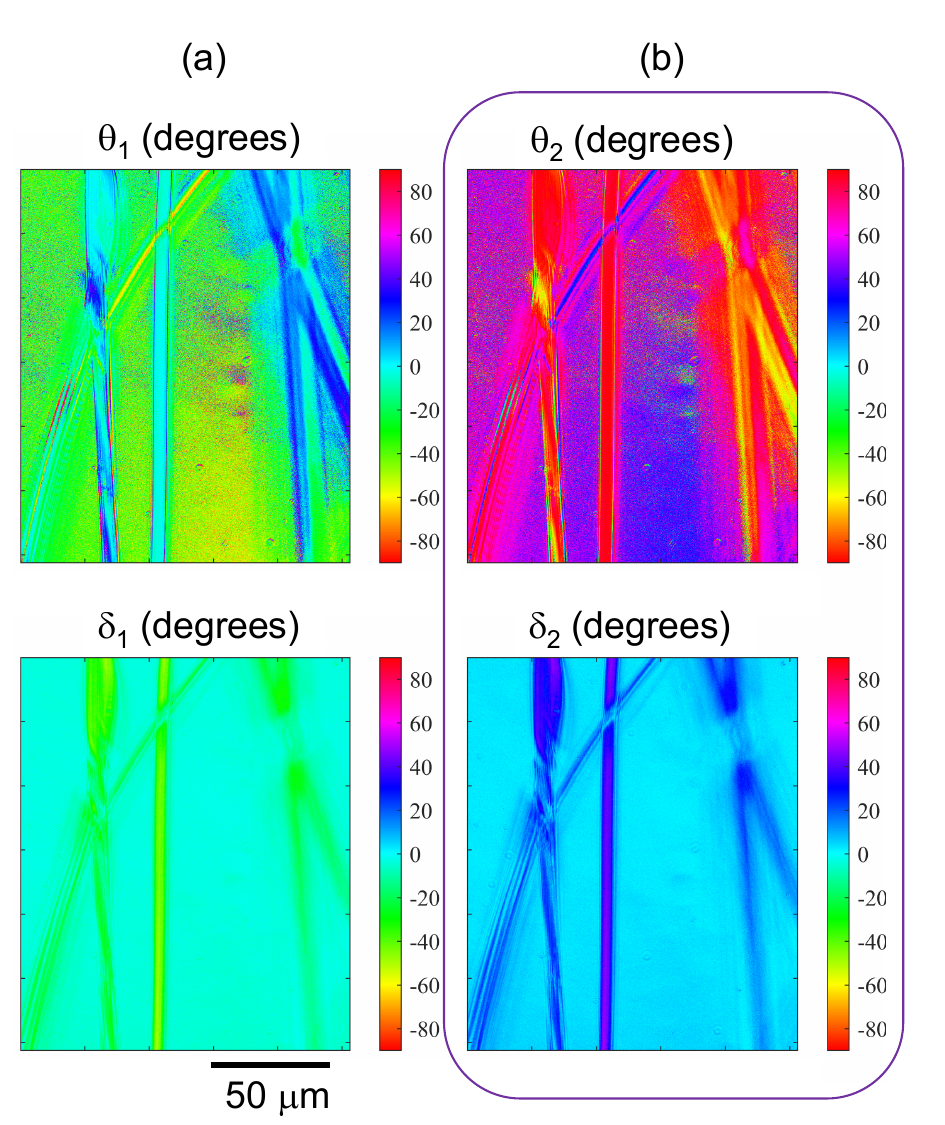}}
\caption{Determination of $(\delta,\theta)$ from analytical solution Eqn.~\ref{e-asol} using experimentally determined $S_{0,1,2}$ from the 4-pol. intensity images. Positive retardance is expected for silk~\cite{18sr17652} and corresponds to the panel (b). 
}\label{f-ambi}
\end{figure}

\begin{figure*}[tb]
\centerline{\includegraphics[width=1\linewidth]{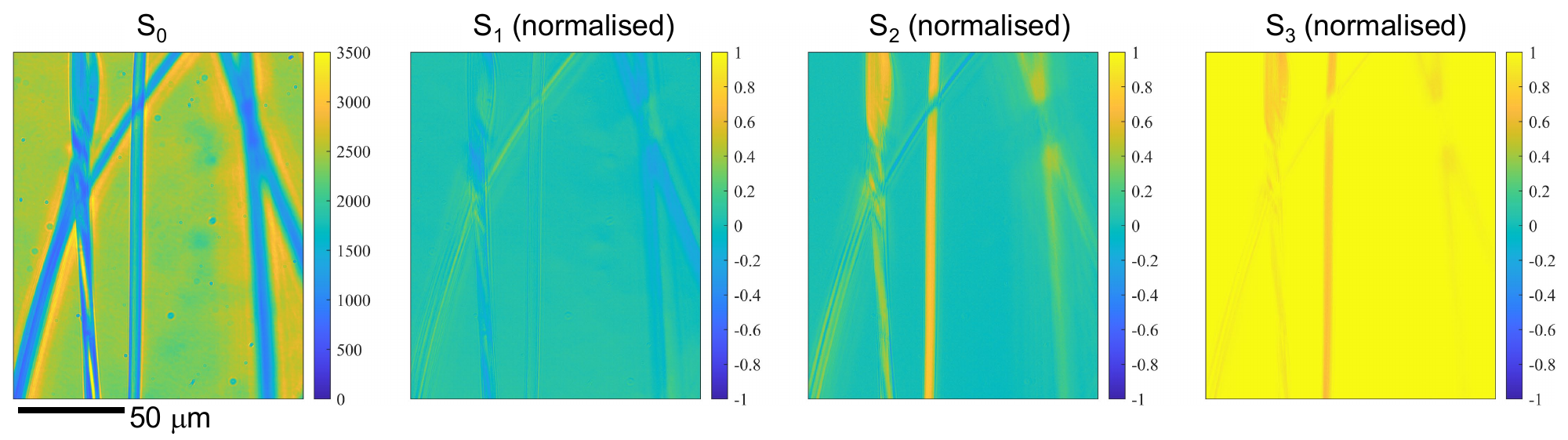}}
\caption{Maps of Stokes' parameters $S_0$ (intensity) and normalised $S_{1,2,3}$ calculated from analytical model Eqn.~\ref{e-asol} (same sample and conditions as in Fig.~\ref{f-ambi}). Objective lens $NA = 0.65$; image taken at $\lambda = 550$~nm using bandpass filter. The $S_3 = \cos\delta$ was calculated from the determined $\delta$; noteworthy, the $S_3 = 0$ when calculated for the entire experimental setup with 4-pol. camera detection, i.e., linear polarisers makes $S_3\equiv 0$. The background has $S_3 = +1$ for the RHC illumination set by the LC $\lambda/4$-waveplate (Fig.~\ref{f-meth}(a)) and would be $S_3 = -1$ for the LHC, since definition of $S_3 = I_{RHC}-I_{LHC}$ is by the intensities at two cross circular polarisations $I_{RHC,LHC}$.  }\label{f-Stok}
\end{figure*}

\begin{figure*}[tb]
\centerline{\includegraphics[width=.85\linewidth]{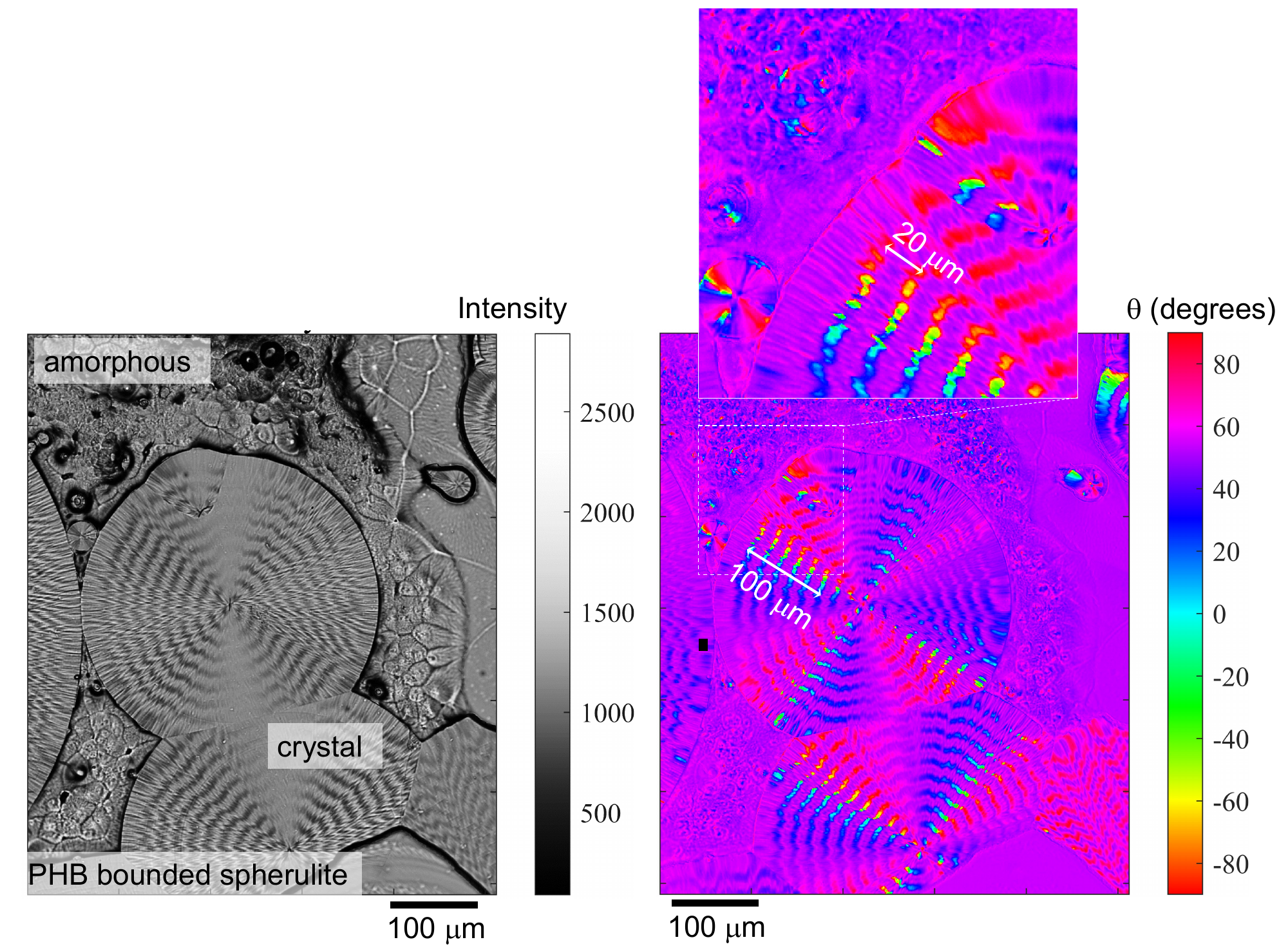}}
\caption{Polyhydroxybutyrate (PHB) bounded spherulite imaged at 550~nm wavelength and the orientation map was calculated via the best fit procedure described here for spider silk. In PHB $n_e < n_o$, hence the best fit was carried out for $-90^\circ\leq\delta\leq 0^\circ$.  Objective lens $10^\times$ with $NA = 0.25$ (Nikon); the resolution $r = 0.61\lambda/NA = 1.34~\mu$m.
}\label{f-phb}
\end{figure*}

Figure~\ref{f-map} shows a summary of the polarisation analysis of a spider silk strand only $\sim 6~\mu$m in diameter using the proposed best fit mode with an additional $\lambda/4$-plate added in front of the sample (retarder). The fit of 4-pol. camera images for each single pixel was carried out using Eqn.~\ref{e-s0}. The retardance $\delta$ and its azimuth $\theta$ inside the silk are retrieved with high fidelity showing a high confidence range of the fit based on residuals $R^2\approx 0.996$ . This high fidelity fit also signifies that there was negligible dichroism and optical activity at the wavelength of illumination, as expected for silk at visible wavelengths. By selecting a region of interest (ROI) on the silk fiber, the maximum retardance $\delta_1 = 47.2^\circ$ 
is obtained for the central cross section of the fiber. The aximuth angle $\theta_1 = 4.6^\circ$ is obtained and corresponds to the slow axis  of silk (along $n_e$ along the fiber).

Figure~\ref{f-cros} shows a cross section measurement of the silk fiber at $\lambda = 550$~nm wavelength. Due to its small diameter, the edges of image are affected by diffraction (imaging was taken with $NA = 0.65$ and the corresponding resolution was  $r = 0.61\lambda/NA = 516$~nm). The two cross sections at max-max intensity and at the baseline varied by only $\sim 4\%$. The retardance $\Delta n d = (\delta\mathrm{^\circ}/360^\circ)\times\lambda = 72.1$~nm, which defined birefringence $\Delta n\equiv n_e-n_o = 0.012$ (for $d = 6.02~\mu$m); for $d = 5.76~\mu$m, $\Delta n = 0.0125$. 

The retardance and its azimuth was determined using the best fit to Eqn.~\ref{e-s0} for a spider silk fiber that was only $\sim 6~\mu$m in diameter. The birefringence of silk $\Delta n \approx 0.012$ was determined. The procedure was accomplished by a single image snapshot using a 4-pol. camera.  

\subsection{All four Stokes parameters from 4-pol. imaging}

A polarisation state is fully determined (a point on Poincare sphere) when all four $S_{0,1,2,3}$ Stokes parameters are known. By addition of a circular polariser (RHC) before the sample (retarder), we showed how retardance $\delta$ and its azimuth $\theta$ can be determined from best fit as described in the previous section. 
The fourth Stokes component $S_3$ can be calculated as $S_3 = \cos\delta$, while the first three are directly determined from the fit of 4-pol. images. 

An analytical method for determination of $(\delta,\theta)$ (Eqn.~\ref{e-asol}) was also used to calculate maps of two equivalent solutions for pairs of $(\delta_1,\theta_1)$ and $(\delta_2,\theta_2)$, which both satisfy the $\tan2\theta = -S'_1/S'_2$ (Fig.~\ref{f-meth}(b)). Figure~\ref{f-ambi} shows results calculated from a pair of experimental values of $S'_1,S'_2$. Solution $(\delta_2,\theta_2)$ (Fig.~\ref{f-ambi}(b)) corresponds to the real case with positive retardance $\delta>0$ values (a negative value of $\delta$ would be expected for the form-birefringent material, which has $\Delta n < 0$, i.e., $n_o > n_e$ ). The ambiguity in $(\delta,\theta)$ is solved by the fact that $n_e > n_o$ in silk and that the slow axis is along the fiber. Another conventional method to solve this issue of fast vs. slow axis is to use a color-shifting plate (usually a 530~nm wavelplate) in cross-polarised imaging, which determines positive and negative contributions to the refractive index with blue-red perpendicular directions corresponding to the $n_e$(slow)-$n_o$(fast) orientations of silk~\cite{23n1894}.

Finally, with $\delta,\theta$ determined via fit or from analytical expression using single a image from a 4-pol. camera, all four Stokes parameters can be defined: $S_0$ (intensity) and $S_{1,2,3}$. Such calculations using an analytical model and a resolution of the uncertainty of the fast-slow axis orientation (Fig.~\ref{f-ambi}(b)) is shown in Fig.~\ref{f-Stok}. If there is no knowledge of the slow/fast axis orientation at the wavelength of interest, both maps $(\delta_{1,2},\theta_{1,2})$ can be calculated from the same data without the need to retake an image (Fig.~\ref{f-ambi}). It is noteworthy that $S_3(\delta)$ is the same when calculated for the two pairs of $(\delta_{1,2}, \theta_{1,2})$ based on experimentally measured $(S'_1,S'_2)$. 

The main advantage of the above presented procedures to determine $\delta$ and $\theta$ is due to inherent capability to calculate them from a single image from a 4-pol. camera. For example, formation of a crystalline phase of organic crystal polyhydroxybutyrate (PHB) from a cooled amorphous melt ($150^\circ$C) takes place with a typical front speed of few micrometers per second for a tens-of-$\mu$m thick bounded spherulite between two transparent plates~\cite{23lpr2200535}. Figure~\ref{f-phb} shows an optical image (intensity) and orientation angle $\theta$ maps obtained by the best fit. This example shows a complex spiraling crystalline pattern of PHB with the spiraling period of $\sim 20~\mu$m revealed via the orientation azimuth. Future studies will be focused on temporal evolution of phase transitions, which can also be imaged at different wavelengths, e.g., at the IR fingerprinting spectral range. The order-disorder front and orientation anisotropy of chemical bonding could be monitored in real time using the two proposed methods based on the fit and via analytical expression using experimentally measured intensities from 4-pol. camera images.

\section{Conclusions and Outlook}

Orientation azimuth of retardance and its value can be extracted without any prior knowledge of orientation from a 4-pol. image using polarised illumination of the sample, addition of a $\lambda/4$ waveplate, and numerical analysis presented in this study. Importantly, this is can be achieved in a single measurement snap-shot with a 4-pol. camera. With knowledge of the retardance $\delta$, all four Stokes parameters can be calculated, i.e., full characterisation of the light state on a detector is obtained and can be linked to the local retardance and its orientation in the sample. This opens new opportunities for imaging of fast-changing events, e.g., phase changes in crystals under microscopy observation at IR or visible wavelengths as well as polarisation analysis of satellite images where natural illumination has prevalent polarisation in the plane of scattering, while a 4-pol. camera is used for imaging. The single image technique presented here for 
Stokes polarimetry is expected to find a number of potential applications, also, at different spectral ranges of electromagnetic radiation, e.g., as fibular structure of bamboo shows linear anisotropy at THz spectral range~\cite{bamboo}. \blue{Since Stokes vector is a 2D projection, the variations inside the sample affect the possibility to decouple a retarded phase due to birefringence from that due to inhomogeneities of thickness or composition. However, for many practical applications where micro-tomed thin slices are imaged, the proposed method should prove useful. If required, true 3D techniques could be applied as recently demonstrated~\cite{Saba:21,Park}.  }    

\small\begin{acknowledgments}
Funding via the ARC Linkage LP190100505 project is acknowledged by S.J.; J.M. and M.R. were funded by JSPS KAKENHI Grant No.~22H02137 and JST CREST Grant No.~JPMJCR19I3. M.R. was funded by JSPS KAKENHI Grant No.~22K14200.
\end{acknowledgments}


\section*{Data Availability Statement}
Data can be made available upon a reasonable request. 
\section*{Statement on Conflict of Interests}
Authors declare no conflict of interests. 


\bibliography{aipsamp}
\appendix
\section{Supplement}
\setcounter{figure}{0}
\makeatletter 
\renewcommand{\thefigure}{S\arabic{figure}}

\begin{figure}[h]
\centerline{\includegraphics[width=1\linewidth]{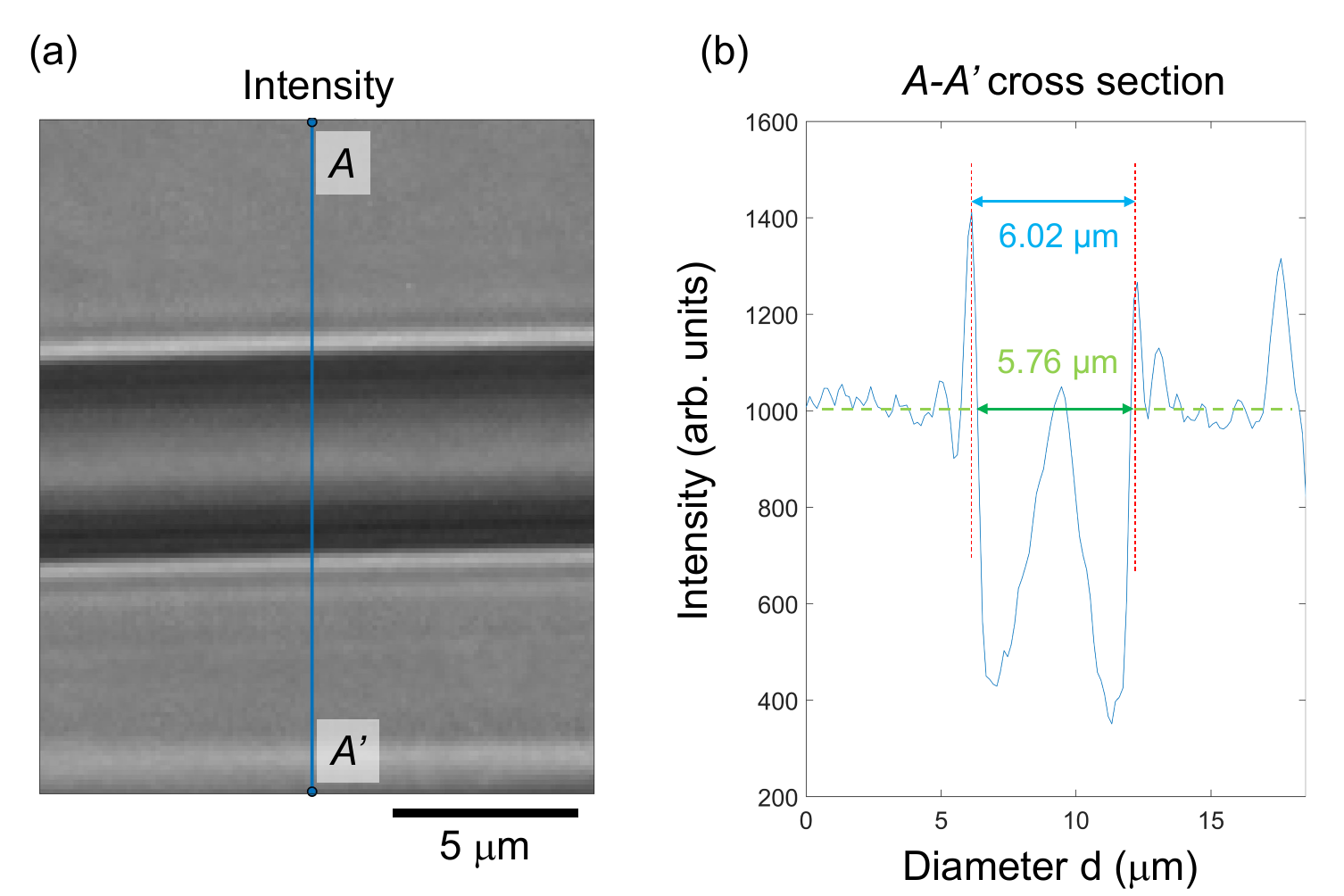}}
\caption{Image (a) and cross section (b) of the yellow spider silk  taken with a $550\pm 10$~nm bandpass filter. The maximum intensity cross section is $4.3\%$ larger than the baseline based image cross section. Same sample and imaging conditions as in Fig.~\ref{f-map}. }\label{f-cros}
\end{figure}

Figure~\ref{f-cros} shows the method used to define the cross section of the silk fiber. The larger cross section from the brightest locations across the fiber were selected for the estimation of birefringence from retardance $\Delta n\times d$. This is a more conservative estimate and the difference from the smaller cross section, as defined by the diameter at the baseline (background intensity), is only $\sim 4\%$ larger.

\begin{figure}[tb!]
\centerline{\includegraphics[width=.5\linewidth]{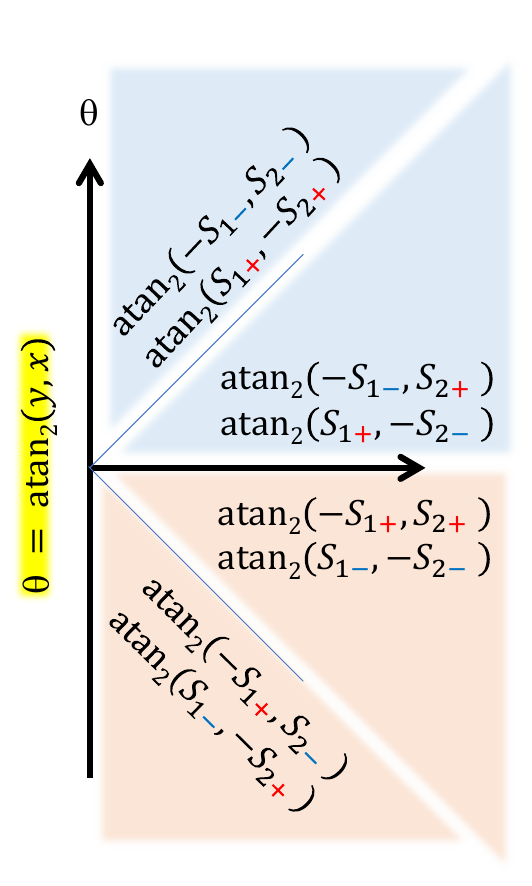}}
\caption{Angular ranges marked for calculation of the azimuth angle $\theta$ using $\arctan_2$ function from the experimentally measured Stokes parameters $S_1$ and $S_2$.  The sub-script sign $S_{1+}$ indicates that $S_1 > 0$ is positive and vice versa for the $S_{1-} < 0$. } \label{f-atan}
\end{figure}

There are four possible combinations of the signs of $S_1$ and $S_2$:  ($S_{1+}$, $S_{2+}$), ($S_{1+}$, $S_{2-}$), ($S_{1-}$, $S_{2+}$), ($S_{1-}$, $S_{2-}$) shown schematically in Fig.~\ref{f-atan}. In addition, $\theta = \arctan_2{(-S_1/S_2)}$, so there are two possible combinations:  $\theta_1 = \arctan_2{(-S_1,S_2)}$ and $\theta_2 = \arctan_2{(S_1, -S_2)}$, but the resulting $\theta$ will be the same for the cases of $-90^\circ<\theta_1<-45^\circ$ and $-45^\circ<\theta_1<0^\circ$ or $0^\circ<\theta_2<45^\circ$ and $45^\circ<\theta_2<90^\circ$.
If $S_1$ and $S_2$ are known and the light is perfectly polarized, we can find $S_3$, since $S_1^2 + S_2^2 + S_3^2 = S_0^2$. However, without use of the circular left and right polarization components to measure Stokes parameters,  $S_3 = 0$ (see, Eqn.~\ref{e-mat}). Also, if $S_3 = \sqrt{S_0^2-S_1^2-S_2^2}$, the information of the circular pol. right-hand or left-hand  is lost.

\begin{figure*}[!h]
\centerline{\includegraphics[width=1.\linewidth]{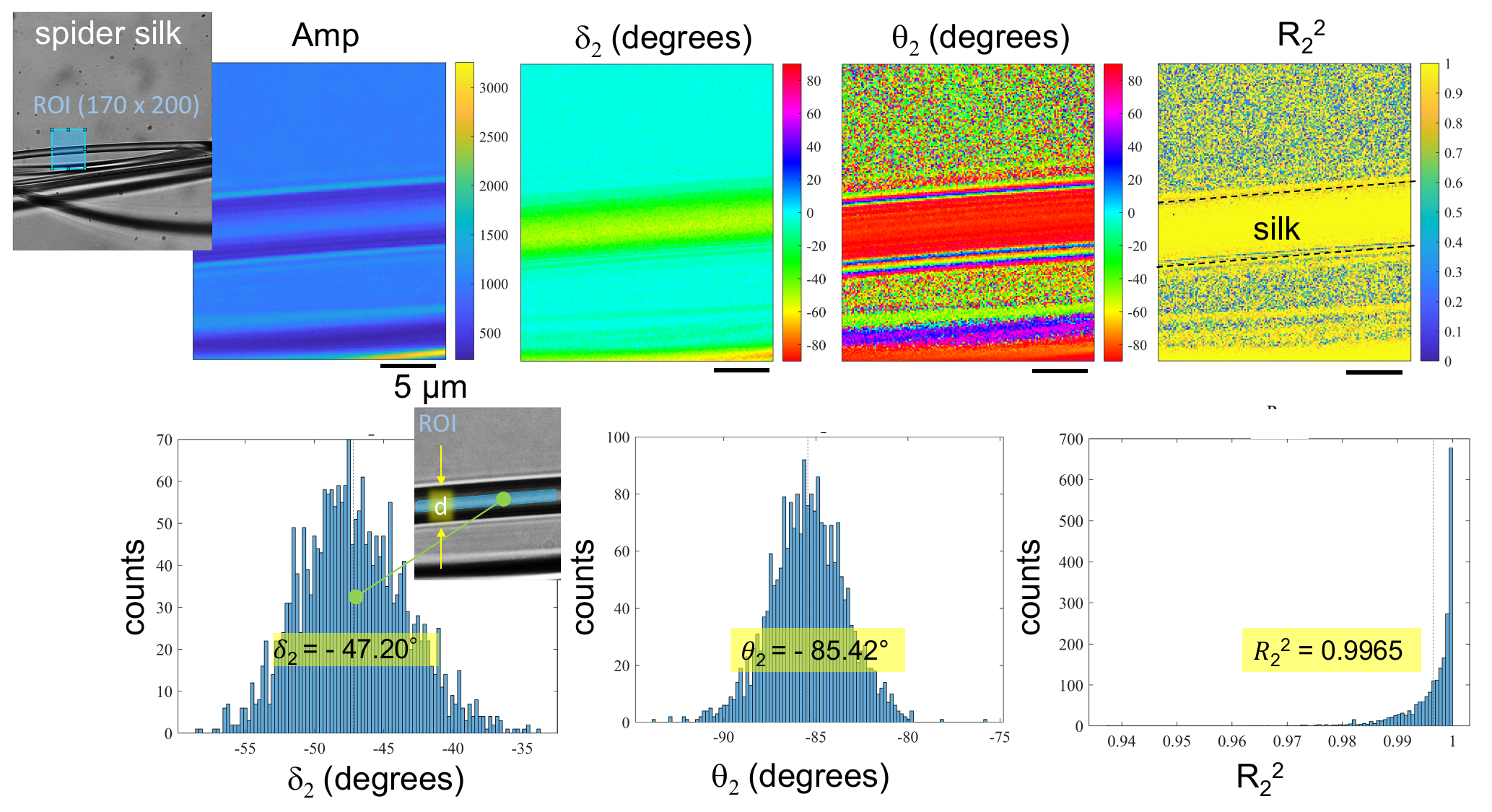}}
\caption{ Same analysis as for Fig.~\ref{f-map} only for the limits of $-90^\circ\leq \delta\leq0^\circ$ and $-90^\circ\leq \theta\leq 90^\circ$.
}\label{f-comp}
\end{figure*}

The second pair of solutions to the best fit $(\delta_2,\theta_2)$ is shown in Fig.~\ref{f-comp}. It is complimentary pair to the $(\delta_1,\theta_1)$ shown in Fig.~\ref{f-map}. As predicted from the analytical solution, $\delta_1 = -\delta_2$ and $\theta_2 = \theta_1-\pi/2$.

\clearpage

\end{document}